\documentclass[3p,times]{elsarticle}

\usepackage{ecrc}

\usepackage{epstopdf}

\volume{00}

\firstpage{1}

\journalname{Nuclear Physics A}

\jid{nupha}

\jnltitlelogo{Nuclear Physics A}

\usepackage{graphicx}
\usepackage{amsmath,amssymb}

\RequirePackage{lineno}

\usepackage{color}

\newcommand{\chka}{\textrm{K}}
\newcommand{\chpr}{\textrm{p}}
\newcommand{\pt}{p_\textrm{T}}
\newcommand{\ptpar}{p_\textrm{T}^\textrm{track}}
\newcommand{\ptjet}{p_\textrm{T, jet}^\textrm{ch}}
\newcommand{\gev}{\textrm{GeV}/c}
\newcommand{\zch}{z^\textrm{ch}}
\newcommand{\dedx}{\textrm{d}E/\textrm{d}x}
\newcommand{\spar}{\boldsymbol{\theta}}
\newcommand{\fra}{f_{ki}}

\usepackage{wasysym}

\begin{document}

\begin{frontmatter}



\title{Measurement of hadron composition in charged jets from pp collisions with the ALICE experiment}

\author{Xianguo Lu (for the ALICE\fnref{col1} Collaboration)}
\fntext[col1] {A list of members of the ALICE Collaboration and acknowledgements can be found at the end of this issue.}
\address{Physikalisches Institut, Ruprecht-Karls-Universit\"at Heidelberg, Germany}




\begin{abstract}
We report the first measurement of charged pion, kaon and (anti-)proton production in jets from hadron colliders. The measurement was carried out with the ALICE detector using $2\times10^8$ minimum bias pp collisions at a centre-of-mass energy of~$\sqrt{s}=7$~TeV at the LHC. We present the $\pi$, K and p transverse momentum ($p_\textrm{T}$) spectra, as well as the spectra of the reduced momentum~(${z^\textrm{ch}\equiv p_\textrm{T}^\textrm{track}/p_\textrm{T, jet}^\textrm{ch}}$), in charged jets of $p_\textrm{T}$ between 5--20~$\textrm{GeV}/c$. The measurement is compared to Monte Carlo calculations. 

\end{abstract}


\end{frontmatter}



\section{Introduction}
\label{sec:intro}

Jets are phenomenological objects constructed to represent partons originating from hard scattering processes. Recently,  neutral  fragments in jets have been studied by the CDF Collaboration~\cite{cdf}. 
However, knowledge of jet fragmentation to charged hadrons currently is only constrained by data from~$\mathrm{e}^+\mathrm{e}^-$ colliders~\cite{Lees:2013rqd,Leitgab:2013qh}. 
The particle identification (PID) capabilities of the ALICE experiment~\cite{Aam08}  at the LHC with sophisticated statistical techniques allow to identify  charged pions, kaons and (anti-)protons in jets  at hadron colliders for the first time. 

The Time Projection Chamber (TPC)~\cite{Alm10} is the main  PID device of ALICE with 
 the specific energy loss ($\dedx$) resolution   in $\chpr\chpr$ collisions of~$\textrm{5--6\%}$.  
The measurement presented here follows the same analysis strategy as the previous ALICE measurement of the inclusive charged particle production in charged jets~\cite{chjet} and explores in addition the TPC performance limit for the challenging PID  in jets.

\section{Analysis details}\label{sec:details}

This analysis is based on a minimum bias (MB) data sample of $2\times10^8$ $\chpr\chpr$ collisions   at  a centre-of-mass energy of $\sqrt{s}=7$~TeV collected with the ALICE detector.  Charged tracks of transverse momentum~${\ptpar>0.15~\gev}$ in the pseudo-rapidity range~$|\eta^\textrm{track}|<0.9$ are reconstructed with the ALICE Inner Tracking System (ITS) and the TPC. They are then grouped into charged jets  with the anti-$k_\textrm{T}$ algorithm  implemented in the FastJet~\cite{Cacciari:2011ma} package and with a resolution parameter $R=0.4$.  
Charged jets  of  transverse momentum $\ptjet$ in three regions~5--10,~10--15,~15--20~$\gev$ are studied separately. The uncorrected differential yields of $\pi$,~$\chka$ and~$\chpr$ in the jets are extracted using the TPC $\dedx$ information. The results are  corrected for detectors effects, including tracking efficiency, track~$\pt$ resolution, jet~$\pt$ resolution, secondary particle and muon (fake pion) contamination.  These  corrections   are based on the  inclusive measurement  in~\cite{chjet} taking into account  particle species dependent effects. The following discussions focus on  the method of  the PID in jets.

The uncorrected differential yields of $\pi$,~$\chka$ and $\chpr$ are provided by the TPC Coherent Fit~\cite{tcf} which is a  fitting program analysing the two-dimensional distribution of the TPC~$\dedx$ signal vs. particle momentum ($p$). It determines  the raw differential particle yields in~$\pt$ from~0.15~$\gev$ to above 20~$\gev$  with an accuracy on average better than~10\% for the 
 ALICE data in~$\chpr\chpr$ collisions. This method is based on the observation that the  mean and width of the~$\dedx$ distribution as well as the fractions of different particle species are all continuous functions of the particle momentum. Using models of  $\dedx$ distributions, denoted as 
$s_k\left(p;\spar\right)$ with a set of \textit{a priori} unknown parameters~$\spar$, for  particle species~$k$ as a function of the particle momentum, a log-likelihood function $l$ is built with additional parameters~$\fra$ for the  fraction of   species $k$ at  momentum bin $i$. The expression for $l$ reads 
\begin{align}
l=\sum_i l_i\left[\sum_{k} \fra s_k\left(p_i;\spar\right)\right],\label{eq:lstat}
\end{align}
where $l_i$ is the log-likelihood function  for the  momentum bin $i$.
 The functional form of the particle fraction $\fra$ is not known, but a continuity condition in the particle momentum can be imposed by adding a regularisation term to Eq.~\ref{eq:lstat} such that~$\fra$ is only allowed to deviate statistically from the interpolated value from neighbouring momentum bins
\begin{align}
l=\sum_i l_i\left[\sum_{k} \fra s_k\left(p_i;\spar\right)\right]+\sum_{k,i}l^\textrm{reg}_{ki}\left(\fra\right),\label{eq:lfull}
\end{align}
where  the regularisation strength  for each particle species and momentum bin contributes equally\footnote{The regularisation term $l^\textrm{reg}_{ki}$ is derived from a Gaussian likelihood and therefore has a proper statistical interpretation.}. The parameters~$\spar$ and~$\fra$ are determined using the maximum likelihood estimation with the full log-likelihood function Eq.~\ref{eq:lfull} from the~$\dedx$ vs.~$p$ distribution. This single optimisation procedure drives \emph{coherently} the full-range constraint on the~$\dedx$ models as well as the constraint on the particle fractions, performing the calibration of~$\dedx$ and the particle yield extraction simultaneously. 

\begin{figure}[t]
\begin{center}
{\includegraphics[height=0.3\linewidth]{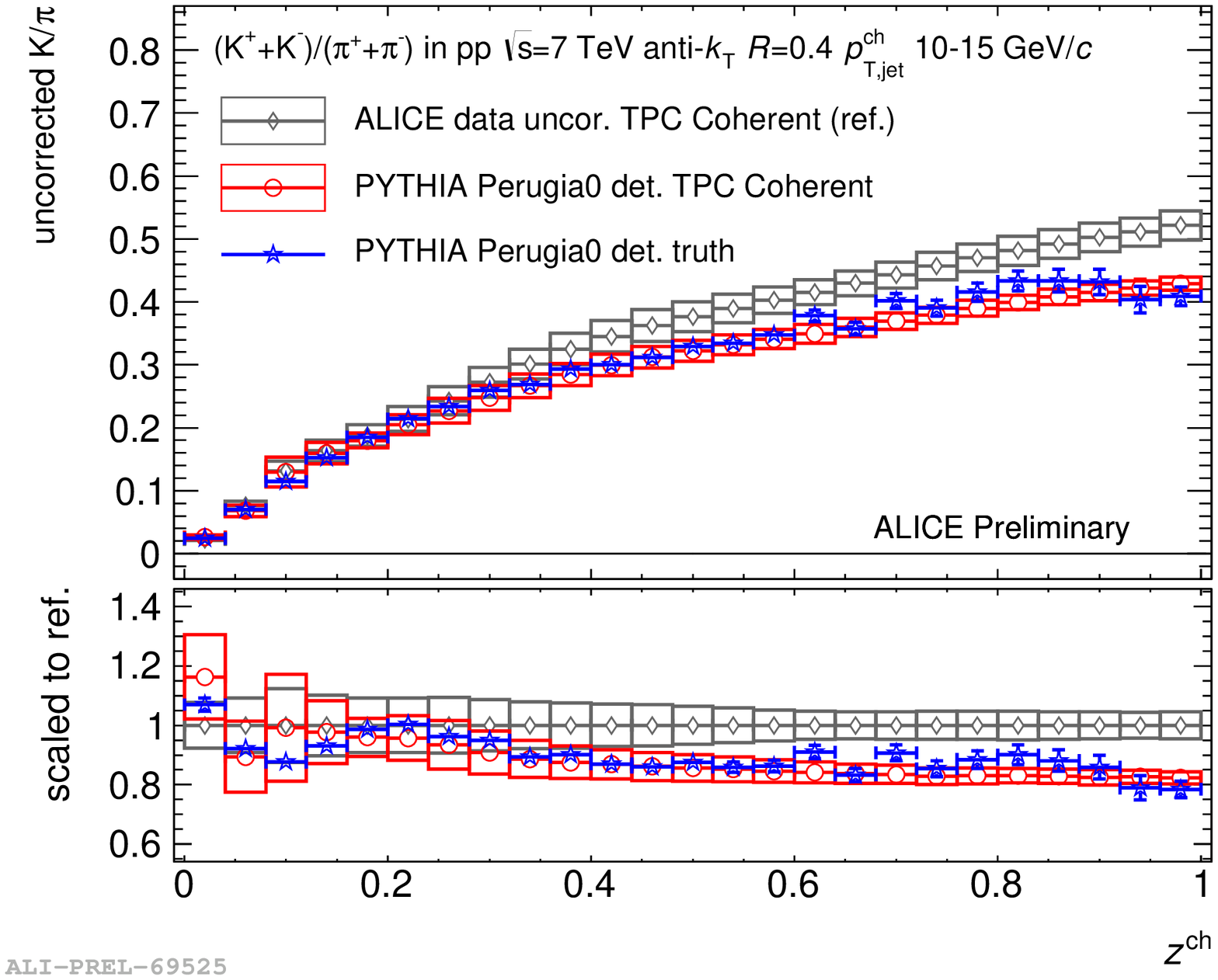}}
{\includegraphics[height=0.3\linewidth]{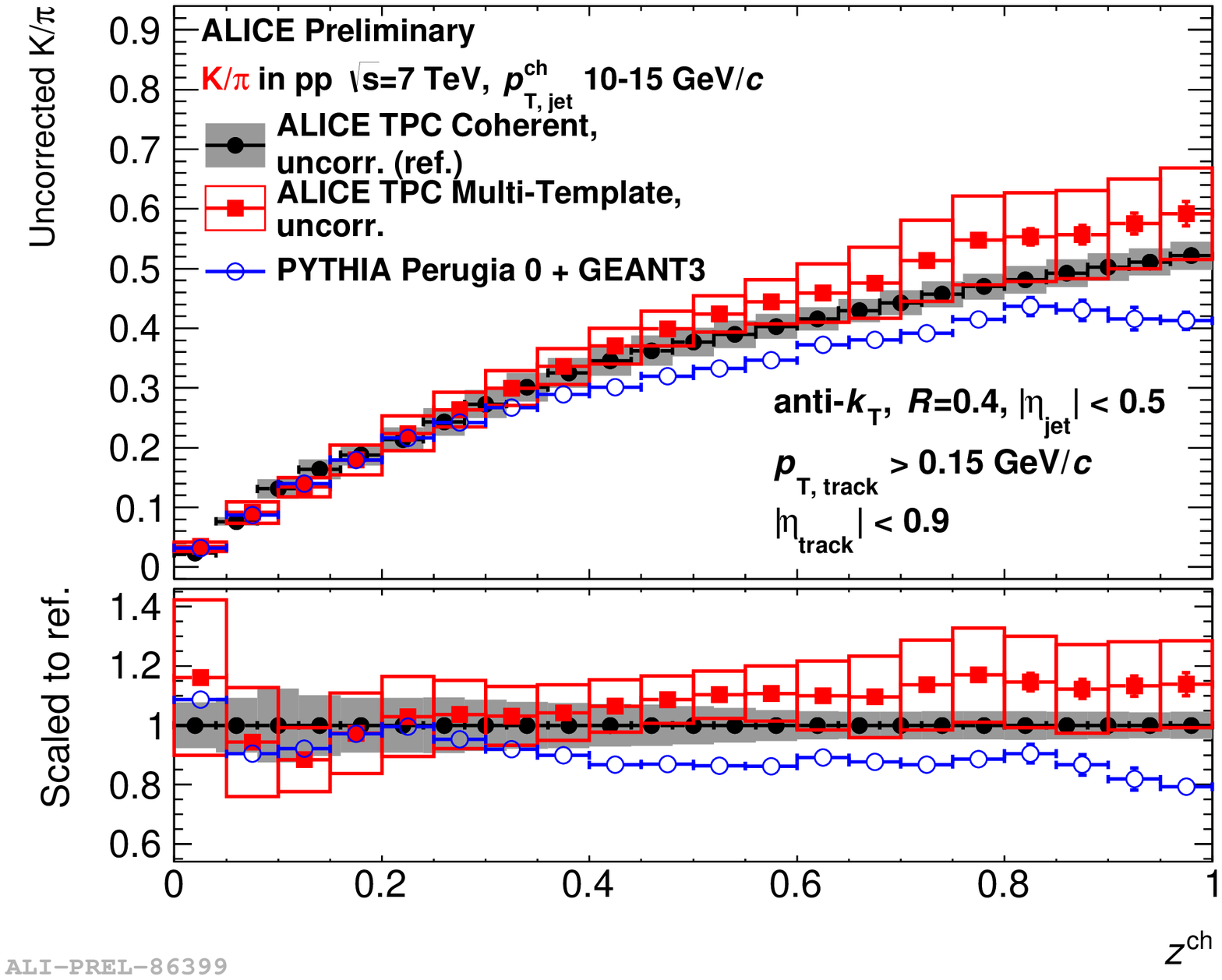}}
\caption{Uncorrected $\chka/\pi$ ratio for charged jet constituents as a function of $\zch\equiv\ptpar/\ptjet$ from data and MC (PYTHIA Perugia0).  \emph{Left}: uncorrected results from data and MC  by the TPC Coherent Fit  compared to the detector level MC truth.  \emph{Right}: uncorrected results from data by the TPC Coherent Fit and the TPC Multi-Template Fit  compared to the detector level MC truth. The error bars are statistical errors and the error boxes represent  systematic errors.}\label{fig:uncor}
\end{center}
\end{figure}

The challenge in  PID using $\dedx$ at high particle momentum ($p>4~\gev$) is that the particle fractions are very sensitive to the mean of the $\dedx$ distribution. For the ALICE $\chpr\chpr$ collision data, the~$\chka$-$\chpr$ separation is about~5\% at high momentum, and consequently a~1\permil~$\dedx$ bias can induce a~2\% bias in the particle fraction. Extensive examination of the systematics in the TPC Coherent Fit has therefore been performed, including  the $\dedx$ model uncertainties,  the stability against the change of the $\dedx$ quality,  the possible particle species dependence of~$\dedx$  (which turns out to be negligible), and mostly critically,  the jet~$\pt$ dependence of~$\dedx$ which is caused by an enhanced track density in jets\footnote{This effect is quantitatively verified from an independent study using electron-positron pairs from $\gamma$-conversions and a realistic track density profile in jets provided by PYTHIA~\cite{Sjostrand:2006za}.}. The last source of systematics is specific to this analysis with respect  to the general case of the identified particle production in inclusive MB data sample. The TPC Coherent Fit performs the~$\dedx$ calibration for different $\ptjet$ data samples separately and  it was found that in these jet $\pt$ regions the $\dedx$ increases by~0.3\% per~5~$\gev$~$\ptjet$ increase. Because with the TPC Coherent Fit the $\dedx$ parameters are automatically adjusted for each~$\ptjet$ sample, no error arises from this effect. As a further verification of the systematic errors, the method is applied in an identical way  also to Monte Carlo (MC) samples (PYTHIA~\cite{Sjostrand:2006za} Perugia0~\cite{Skands:2010ak} tune) with the full ALICE detector simulation and reconstruction. As is shown in Fig.~\ref{fig:uncor}~(\emph{left}), the uncorrected $\chka/\pi$ ratio extracted by the TPC Coherent Fit from the MC sample  is consistent with the detector level MC truth. In the same figure, the uncorrected  ratio from data is also shown; it can be seen that MC deviates from the data beyond the systematic errors. 

The TPC Coherent Fit is further cross-checked with an independent method --  the TPC Multi-Template Fit, which generates templates of TPC $\dedx$ distributions with the particle fractions as the only unknown parameters after  predetermination of the $\dedx$ distributions. In this method,  the $\dedx$ response up to  intermediate momentum is determined in detail with pure particles  selected from MB data samples via the TPC $\dedx$, the  Time-Of-Flight (TOF) signal and the decay products of  $\chka^0_\textrm{S}$, $\Lambda$ and $\gamma$. The response  at  high momentum  is determined by the  $\dedx$ model fitting to the pure samples; the particle fractions are estimated by minimising the difference between the template and the data. The two methods have different sources of systematic uncertainties and, as  shown in  Fig.~\ref{fig:uncor} (\emph{right}), the results by both methods are consistent within the estimated systematic errors.


\section{Results}\label{sec:res}

\begin{figure}[t]
\vspace{-0.5cm}
\begin{center}
{\includegraphics[width=0.98\linewidth]{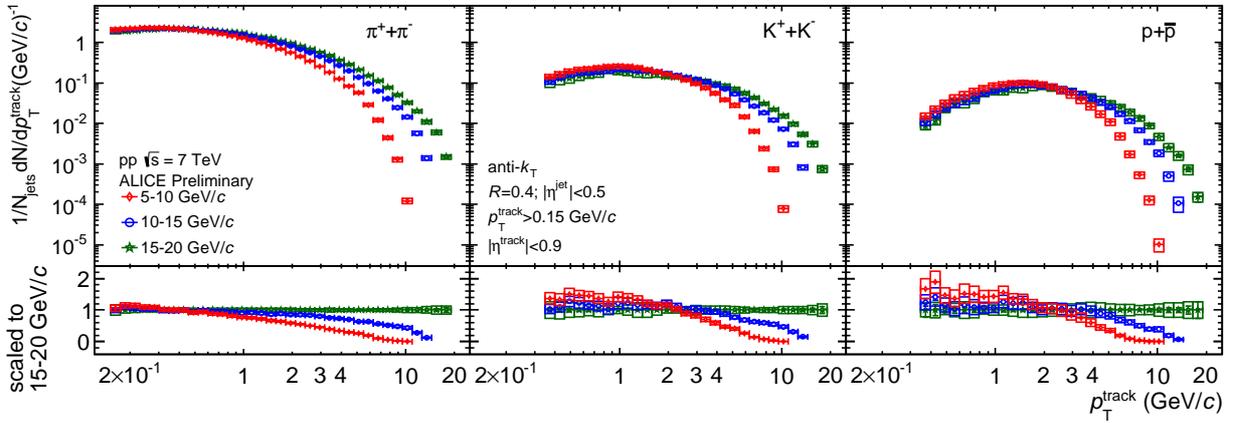}}
\caption{Corrected $\pt$-spectra of $\pi$ (\emph{left}), $\chka$ (\emph{middle}) and $\chpr$ (\emph{right}) in charged jets from  $\chpr\chpr$ collisions at~7~TeV. The spectra are shown for three different charged jet~$\pt$ ranges indicated by the legend.} 
\label{fig:yield}
\end{center}
\end{figure}

The fully corrected $\pt$-differential  yields per jet of $\pi$, $\chka$ and $\chpr$ in charged jets are shown in Fig.~\ref{fig:yield} for the three jet~$\pt$ intervals.
They drop over 3--4 orders of magnitude with increasing particle~$\pt$. Larger~$\ptjet$ gives rise to higher  yields for high $\pt$ particles, the ordering being reversed at around~0.4~(2)~$\gev$ for $\pi$ ($\chka$ and~$\chpr$). 

\begin{figure}[t]
\begin{center}
{\includegraphics[width=0.68\linewidth]{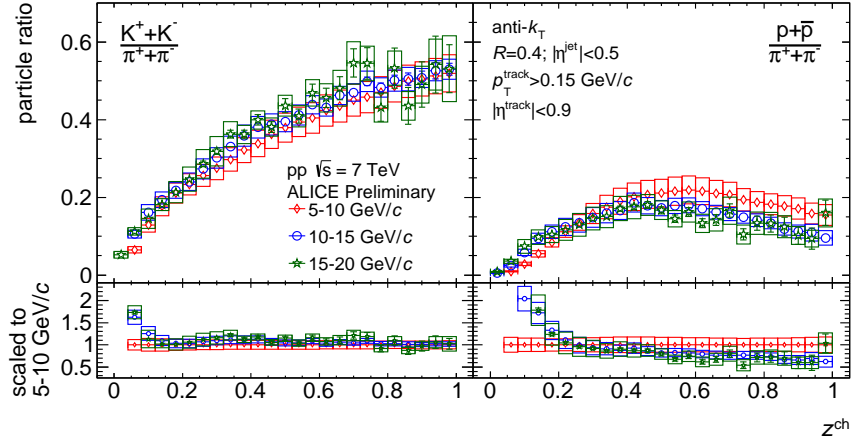}}
\caption{$\chka/\pi$ and $\chpr/\pi$ ratios as a function of $\zch$  for charged jet constituents from $\chpr\chpr$ collisions at~7~TeV.}
\label{fig:ratio}
\end{center}
\end{figure}

The particle ratios $\chka/\pi$ and $\chpr/\pi$ in charged jets as a function of the reduced momentum $\zch\equiv\ptpar/\ptjet$ are shown in Fig.~\ref{fig:ratio}.  The monotonic increase of the $\chka/\pi$ ratio  to 0.5--0.6   shows that the strangeness fraction in jets increases with~$\zch$, while the maxima of the $\chpr/\pi$ ratio at~$0.5<\zch<0.6$ indicate that  leading baryon production in jets is suppressed. Comparing the particle ratios for different jet $\pt$, we observe a $\zch$-scaling at $\zch>0.2$ for $\chka/\pi$ among~$\ptjet$~5--10, 10--15, 15--20 $\gev$ and for $\chpr/\pi$ between 10--15 and~15--20~$\gev$.

\begin{figure}[t]
\vspace{-0.5cm}
\begin{center}
{\includegraphics[width=0.98\linewidth]{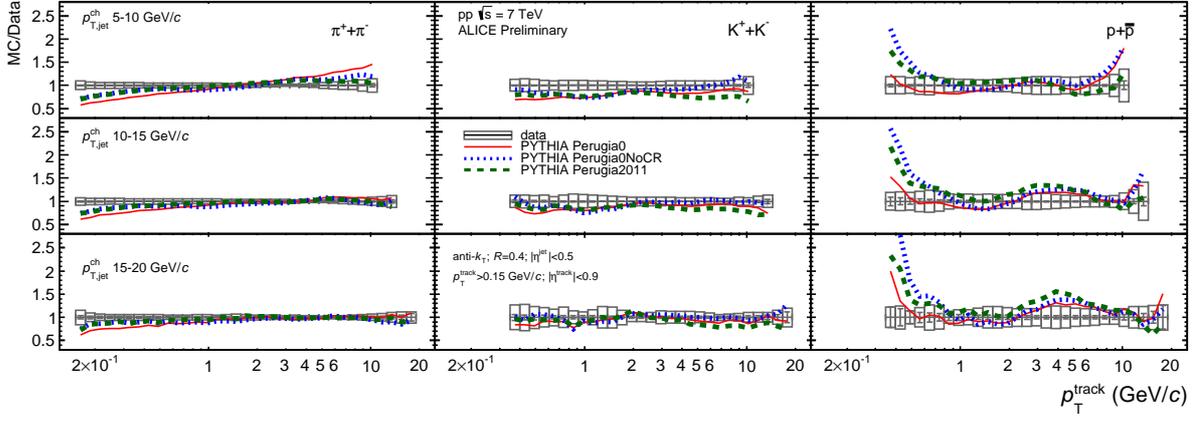}}
\caption{Data-PYTHIA comparison for the particle $\pt$-spectra in charged jets from  $\chpr\chpr$ collisions at~7~TeV.}\label{fig:yieldmc}
\end{center}
\end{figure}

  The results are compared  in Fig.~\ref{fig:yieldmc} to three sets of PYTHIA~\cite{Sjostrand:2006za} tunes, Perugia0, Perugia0NoCR  (\emph{CR} stands for  \emph{colour reconnection}) and Perugia2011~\cite{Skands:2010ak}. PYTHIA describes the~$\pt$-spectra better at large~$\ptjet$ for high~$\pt$ particles. The best agreement is obtained for kaons while PYTHIA undershoots (overshoots) low $\pt$ pions (protons). The data are reproduced in general within~30\% accuracy except for the low~$\pt$ protons for which the deviation goes up beyond~100\% at around $0.4~\gev$. In addition, PYTHIA reproduces the  maxima of the proton spectra but fails to describe the width and the high $\pt$ slope. Furthermore, of
the tunes tested here, Perugia0NoCR gives the best description of the kaon
spectra at~$\ptjet$ 5--10 and~10--15~$\gev$.



\section{Conclusions}\label{sec:conc}

We have presented the first measurement of  jet fragmentation to charged hadrons at hadron colliders from ALICE. The yields of  $\pi$,~$\chka$ and~$\chpr$ in charged jets from $\chpr\chpr$ collisions are extracted with advanced PID techniques. The results show that 
the strangeness fraction increases with~$\zch$ while the leading baryon production is suppressed at high~$\zch$. PYTHIA simulations are able to reproduce the data within 30\% accuracy with increasing tension at low $\pt$.








\end{document}